\documentclass[ reprint, amsmath, amssymb, aps, prd]{revtex4-1}

\newcommand\Tstrut{\rule{0pt}{2.6ex}}       % "top" strut
\newcommand\Bstrut{\rule[-0.9ex]{0pt}{0pt}} % "bottom" strut
\newcommand{\TBstrut}{\Tstrut\Bstrut} % top&bottom struts

\usepackage[english]{babel}
\usepackage{amsbsy}
\usepackage{amstext}
\usepackage{amsmath}  
\usepackage{amssymb}
\usepackage{graphicx}
\usepackage{gensymb}

\usepackage{pst-node} % \left<- for color test
\usepackage{diagbox}
\usepackage{multirow}
\usepackage{todonotes}

\usepackage{float}
\makeatletter
\def\captionof#1#2{{\def\@captype{#1}#2}}
\makeatother
\begin{document}
\title[something]{Measurement of sub-dominant harmonic modes for gravitational wave emission from a population of binary black holes}
\author{ B. D. O'Brien$^{1}$, C. F. Da Silva Costa$^{1}$,  S. Klimenko$^{1}$}
\address{(1) University of Florida, Gainesville, Florida, USA }\email{brendandobrien@ufl.edu}
\date{\today}

\begin{abstract}{ 

    Measurements of multiple harmonic modes in the gravitational wave signals from binary black hole events could provide an accurate test of general relativity, however they have never been observed before. The sub-dominant modes, other than the main $(\ell = 2,m = 2)$ mode, are weak in amplitude and thus difficult to detect in a single event at the current sensitivity of the gravitational wave detectors. To recover sub-dominant modes, we propose an unmodeled method for summation of signals from a population of binary black holes. The method coherently stacks all the signal modes, effectively increasing their signal-to-noise ratio, so the amplified signal can be extracted from the noisy data. To test the method, we consider simulated numerical relativity waveforms including sub-dominant modes up to $(5,5)$.
    %In addition, our parameter space is chosen to emulate the binary black hole events which have been detected through the second observing run of Advanced LIGO.
    We inject simulated signals from a population of binary black holes into data from the first observing run of Advanced LIGO and utilize the coherent WaveBurst algorithm for signal detection and reconstruction. Using only the waveforms reconstructed by coherent WaveBurst, i.e., no \textit{a priori} information about the signal model, we determine the transformation to coherently synchronize the merger and post-merger of one signal to another.
We demonstrate the synchronization of the injected signals and show the efficient stacking of the $(2,2)$ mode and the sub-dominant modes.
}
\end{abstract} 
\maketitle

%#############################################################################################<#####
\section{Introduction}
The first gravitational wave detection, GW150914, proved the existence of binary black hole (BBH) systems and brought with it excitement to the newly found field of gravitational wave astronomy~\cite{bib:GW150914}.
In total, LIGO has detected ten BBH gravitational wave signals, and the current BBH merger rate is estimated to be $52.9_{-27.0}^{+55.6}\,\mathrm{Gpc}^{-3}\mathrm{yr}^{-1}$~\cite{bib:catalog_paper,bib:catalog_paper_2,bib:GW150914,bib:GW151226,bib:GW170104,bib:GW170608,bib:GW170814}.
The current network of Earth-based detectors consists of LIGO Hanford, LIGO Livingston, and Virgo~\cite{bib:aLIGO,bib:Virgo}.
With these detectors approaching design sensitivity, LIGO could possibly detect hundreds of BBH events in the near future.

Following the discovery of BBH systems, attention now turns to analyzing the data from BBH detections and understanding the underlying physics. Strong gravitational fields produced by merging black holes allow us to search for possible deviations from general relativity (GR) with unprecedented sensitivity and understand the nature of a black hole~\cite{RuffiniWheeler71}.

To unravel the information contained in gravitational wave emission of a BBH system, it is useful to decompose the emission into spin-weighted spherical harmonics through multipole expansion~\cite{bib:AMSsilicium}. For most BBH systems, the gravitational radiation is dominated by the ($\ell = 2$, $m = 2$) mode.
In this paper, all other harmonic modes will be referred to as \textit{sub-dominant modes}.
In general, sub-dominant modes correspond to high frequency GW emission, making it possible to separate these modes from the $(2,2)$ mode.
%(the phrasing \textit{higher order modes} is often used synonymously).
Sub-dominant harmonic modes are present in all three phases of the gravitational wave signal: inspiral, merger, and ringdown. The mode structure is a unique signature of the dynamics of the system evolving in extreme gravitational fields. In the ringdown phase, the spherical harmonics evolve into monochromatic damped sinusoids known as quasinormal modes (QNMs), and these modes 
%It follows from the no-hair Theorem that QNMs
are determined by the nature of the remnant black hole~\cite{bib:damped_osc,bib:Press71,bib:QNM_sound}.
The study of QNMs, known in the literature as \textit{black hole spectroscopy}~\cite{bib:Detweiler,bib:bh_spectroscopy_dreyer,bib:bh_spectroscopy_berti}, can identify Kerr black holes and inform us on possible corrections to GR.

No sub-dominant modes have been observed so far for the detected BBH events.
They are expected to be weak, particularly for equal mass, circular, and non-spinning systems.
It is possible that we could detect sub-dominant modes at increased LIGO sensitivity, but it is unlikely that we could measure QNMs even from high signal-to-noise ratio (SNR) events with Advanced LIGO~\cite{bib:bh_spectroscopy_berti}.
Consequently, current methods for measuring sub-dominant harmonic modes and QNMs from a single BBH event are focused on future generations of ground and space-based detectors~\cite{bib:bh_spectroscopy_berti, bib:caudill, bib:no-hair_TIGER, bib:dain, bib:ringdown_London}. 

Stacking techniques (coherent summation) may solve a technical problem of detection of sub-dominant modes with the current detector sensitivities.
As an example, optimal synchronization and summation of $N$ signals effectively increases the SNR of the combined signal by a factor of ${N}^{1/2}$.
The idea of stacking multiple BBH signals is not new, but it has not yet been completely solved.  
Previous work by Yang~\textit{et al.} focused on synchronizing the $(2,2)$ and $(3,3)$ modes from a population of BBH events by using a model of the inspiral signal to determine the frequency and phase of each mode at the ringdown stage~\cite{Yang_stacking}.
C. F. Da Silva~\textit{et al.} 
synchronized the $(2,2)$ QNM for a population of BBH events by using a model of the QNM frequencies and decay rates~\cite{Filipe_stacking}. 

The motivation of this paper is to introduce a method for detection of sub-dominant modes during the merger and post-merger of BBH systems without the use of \textit{a priori} knowledge about the signal model.
We propose a generic method for constructive synchronization and summation of multiple BBH events by applying a transformation to their GW waveforms consisting of a time rescale, a time shift, a frequency shift, and a phase shift.
This transformation is determined strictly by maximizing the overlap between multiple signals, and therefore makes no assumptions about the astrophysical sources.
To capture the high field dynamics of the BBH signals, we limit our synchronization to a time window from the last cycles of inspiral through the post-merger. 

In Section \ref{sec:simulations}, we detail the parameter space of the simulated waveforms used for this study. In Section \ref{sec:MethodDescription}, we present the coherent summation method for an arbitrary number of BBH events. In Section \ref{sec:Results} we discuss the results of applying the method to 16 BBH signals. In Section \ref{sec:Conclusion}, we discuss the conclusions of the study and briefly describe work to be carried out in the future.

\section{Simulated Waveforms}
\label{sec:simulations}

For this study, we use simulated BBH GW waveforms composed of spherical harmonic modes up to (5,5) from the SXS numerical relativity group~\cite{SXS:catalog,bib:sxs2,bib:sxs3}.
The parameter space for the simulated waveforms used in this study are chosen to emulate the emerging population of BBH events which have been detected by LIGO.
We select 16 BBH waveforms and limit our parameter space to systems with mass ratio $q\leq3$, total mass $70\,\mathrm{M}_{\odot}\leq M_{tot}\leq80\,\mathrm{M}_\odot$, zero initial spin, and zero eccentricity orbits.
Thusfar, GW150914 has the highest SNR for any BBH event.
The randomly selected source distance, sky location, and orientation are confined such that the recovered SNR is less than the SNR of GW150914. 
Exact parameters used for the simulated waveforms in this analysis can be found in Table~\ref{tab:injections}.

\begin{table}[ht]
\caption{
Simulated numerical relativity waveforms which are added to the H1-L1 O1 data.
The BBH events are detected and recovered with the cWB pipeline, and ordered by estimated chirp mass.
The network SNR for detected events is determined with the cWB pipeline.
}
\begin{center}
\begin{tabular}{ c c c c c }
\hline
\hline
Event & $M_c/\mathrm{M}_\odot$ & $M_{tot}/\mathrm{M}_\odot$ &  $q$ & Network SNR \TBstrut\\  \hline 
00 & 42.0 & 76.5 & 1.5  & 22.3  \Tstrut\\
01 & 40.2 & 74.5 & 1.0  & 24.5 \\
%02 & 39.8 & 76.8 & 3.0  & 19.4 \\ 
02 & 38.4 & 75.9 & 2.3  & 20.8 \\
03 & 38.2 & 76.2 & 1.0  & 18.5 \\
04 & 37.8 & 76.2 & 2.3  & 21.3 \\
05 & 37.2 & 76.4 & 1.0  & 24.7 \\
06 & 37.2 & 76.0 & 2.5  & 18.9 \\
07 & 37.0 & 75.7 & 1.0  & 23.4 \\
08 & 36.8 & 76.1 & 2.0  & 17.3 \\
09 & 36.8 & 75.7 & 2.0  & 24.0 \\
10 & 35.0 & 74.8 & 2.3  & 22.9 \\
11 & 34.6 & 76.6 & 2.3  & 20.4 \\
12 & 33.8 & 76.8 & 3.0  & 17.5 \\
13 & 32.8 & 75.9 & 1.5  & 23.0 \\
14 & 30.6 & 75.7 & 3.0  & 20.0 \\ 
15 & 29.8 & 76.7 & 2.0  & 19.4 \Bstrut\\ \hline 
\hline

\end{tabular}
\end{center}
\label{tab:injections}
\end{table}

For this analysis, the Hanford-Livingston detector network (H1-L1) is considered, but the analysis could be easily expanded to any number of detectors.
The simulated BBH waveforms are injected (added) into data from the first observing run O1 of advanced LIGO.
The coherent WaveBurst (cWB) pipeline is utilized for signal detection and reconstruction~\cite{sergey2005,sergey2016}. 
For each detected event, cWB produces a list of estimated parameters and an unmodeled reconstructed waveform in each detector. 

\section{Synchronization Method}
\label{sec:MethodDescription}

To determine the transformation parameters for the synchronization procedure, we exclusively use the cWB waveforms recovered for each BBH event, exploiting no \textit{a priori} information about the signal model.
To simplify the explanation of the stacking procedure, we first introduce the method for $N=2$ signals in sub-section~\ref{sec:Stackingfortwo} and expand this discussion to $N>2$ signals in sub-section~\ref{sec:StackingforMany}.

\begin{figure}[htpb] % htbp stands for "here, top, bottom, page" 
\includegraphics[keepaspectratio=true,width=\linewidth]
{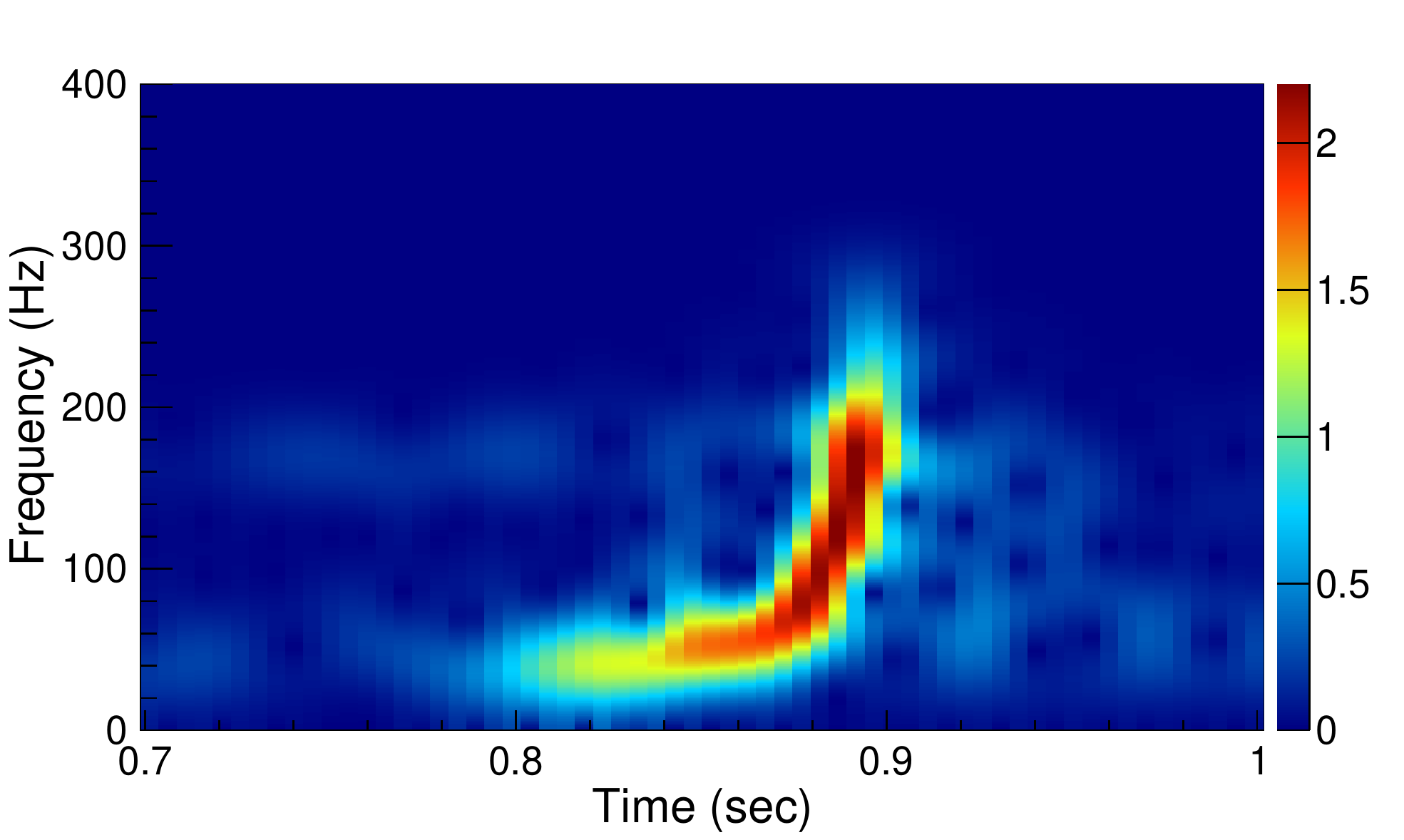}
\caption{
The signal is a numerical relativity waveform added to H1-L1 O1 data (Event 00 from Table~\ref{tab:injections}). Time-frequency representation of the H1 waveform reconstructed by cWB. The transformation applied in this analysis alters the chirp behavior of a given signal.
}
\label{fig:tf_map_test}
\end{figure}

\subsection{Stacking procedure for $N=2$ signals} 
\label{sec:Stackingfortwo}

%The mode structure for a population of BBHs is independent of the GW source sky location $(\theta, \phi)$.
%Since the time delay between detectors is dependent only on the source sky location, it can be artifically eliminated without loss of information to the GW mode structure.
%The time delay between detectors for each BBH event is elminated by applying a timeshift equal to the time of flight, as estimated by cWB, to the reconstruction from one detector stream. We arbitrarily choose to timeshift the reconstruction in L1.

For each BBH event, we consider the waveforms reconstructed by cWB for Hanford ($h_H$) and Livingston ($h_L$). To synchronize two BBH events, we transform the reconstructed waveforms corresponding to one signal ($h_{H1}$ and $h_{L1}$) to match the reconstructed waveforms of the second signal ($h_{H2}$ and $h_{L2}$).
As briefly mentioned earlier, the parameters used to synchronize multiple BBH events are: the time rescale $\alpha$, the time shift $\Delta t$, the frequency shift $\Delta f$, and the phase shifts $\gamma_{H}$, $\gamma_{L}$---corresponding to each of the two detector data streams. Below we refer to this set of parameters as the transformation parameters $\Omega_{12}$.

The mode structure of a BBH event is independent of sky location and detector antenna patterns.
Thus, we choose to simplify the synchronization procedure by introducing a time shift which artifically eliminates the time of flight between the detectors.

The time-frequency representation for a waveform reconstructed by cWB is shown in Figure~\ref{fig:tf_map_test}.
In the time-frequency domain, rescaling a waveform equates to squeezing the waveform along the time axis and stretching the waveform along the frequency axis (or vice-versa). This effectively changes the chirp time-frequency structure of the event.
In addition, the time shift $\Delta t$ and the frequency shift $\Delta f$ correspond to a translation across the time and frequency axes respectively.
Generally speaking, these three parameters---$\alpha$, $\Delta t$, and $\Delta f$---are used to align the chirp curves of different events in the time-frequency domain. 
The same values for these parameters are applied to the Hanford and Livingston reconstructed waveforms ($h_{H1}$ and $h_{L1}$).

\begin{figure*}[htpb] % htbp stands for "here, top, bottom, page" 
\includegraphics[keepaspectratio=true,width=\textwidth]{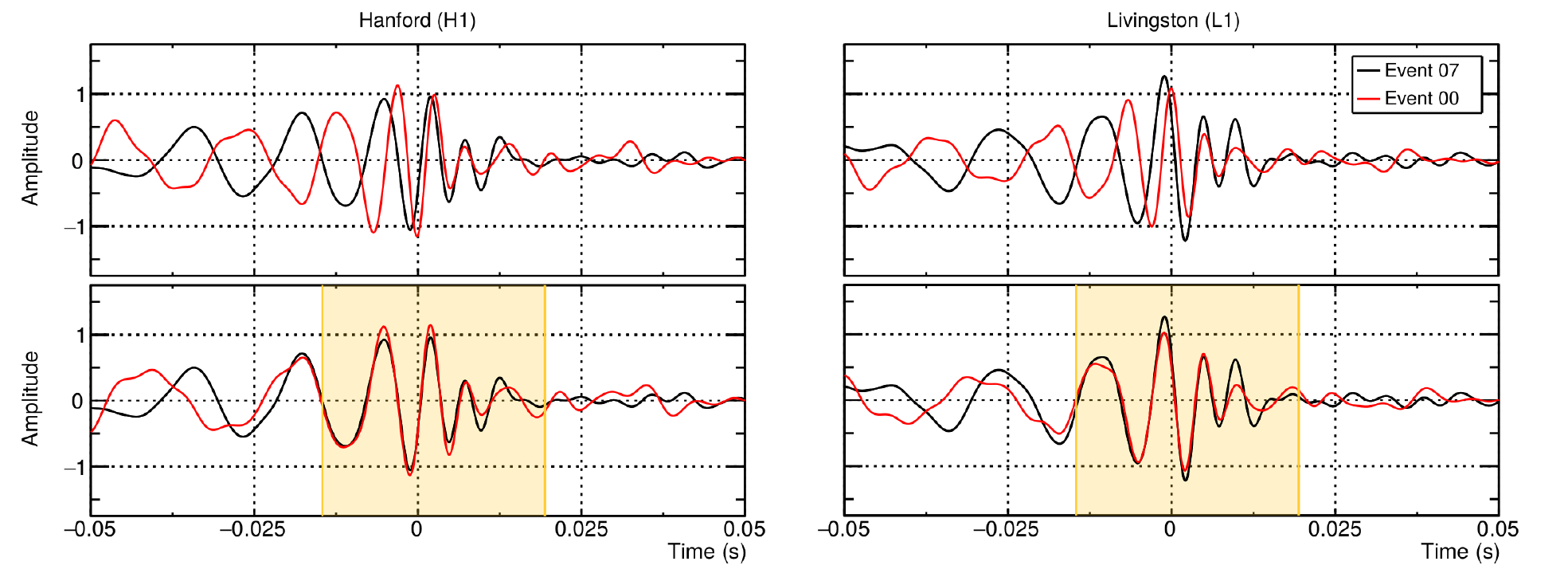}
\caption{
Synchronization method applied to two reconstructed BBH events (simulated events 00 and 07 from Table~\ref{tab:injections}).
Numerical relativity waveforms are injected into H1-L1 data stream, and signals are detected and reconstructed using the cWB pipeline.
\textit{left:} reconstructed H1 waveforms, \textit{right:} reconstructed L1 waveforms. 
\textit{Top row:} Initial synchronization of two signals, a timeshift is applied to align the reference times of the two events: $\tau_1 = \tau_2$.
Time $t=0$ corresponds to the reference time $\tau_1$.
\textit{Bottom row:} The optimal transformation parameters $\Omega_{12}$ are applied to maximize the overlap between the reconstructed waveforms corresponding to the two BBH events.
The time window over which the overlap of the two signals is maximized is depicted in yellow. 
}
\label{fig:stack2}
\end{figure*}

The phase of the waveform reconstructed by cWB is dependent on the sky location and the antenna patterns of the detector. Therefore, it is necessary to apply an independent phase shift for each detector.
We apply the phase shift $\gamma_{H}$ to the Hanford waveform $h_{H1}$ and the phase shift $\gamma_{L}$ to the Livingston waveform $h_{L1}$. 
Omitting the detector index, the phase shifted waveforms are: 
\begin{align}
\label{eq:phase_shift}
\begin{split}
h^{'}_1(t) = h_1(t) cos(\gamma) - H_1(t) sin(\gamma)\, ,
\\
H^{'}_1(t) = H_1(t) cos(\gamma) + h_1(t) sin(\gamma)\, ,
\end{split}
\end{align}
where $H_1$ is the quadrature  waveform, obtained by applying a $90\degree$ phase shift to the reconstructed waveform $h_1$. 
%The frequency shift $\Delta f$ is applied in a similar way as shown in the Appendix.
Before determining the values of the transformation parameters, a reference time is established for each event by using the reconstructed amplitude envelope:
\begin{equation}
A(t) = \sqrt{h^2(t) + H^2(t)}\, .
\end{equation}
The amplitude envelope is determined for both the Hanford waveform ($A_{H}(t)$) and Livingston waveform ($A_{L}(t)$).
We define the reference time $\tau$ for a single BBH event to be the time corresponding to the maximum of the combined ampltiude envelope $\sqrt{A_{H}^2(t) + A_{L}^2(t)}$.
%By adding the amplitude envelope from both detectors $A_{H}(t) + A_{L}(t)$, we define the reference time $\tau$ for a single BBH event to be the time corresponding to the maximum envelope amplitude.
The reference time does not necessarily correspond to an astrophysical occurance, such as the merger time, but is used for the initial synchronization of  multiple signals (see Figure~\ref{fig:stack2}, top panels).
%As a zeroth order approximation to synchronization, we apply a timeshift to align the reference times of two events $\tau_0$ and $\tau_1$ (see Fig.~\ref{fig:stack2}, top panels).
Corrections to this time shift are applied later in the synchronization procedure.

To further synchronize two signals, we introduce a time window within which the values of the transformation parameters are determined to maximize the overlap between the signals' waveforms.
The selected time window, displayed in the bottom panels of Figure~\ref{fig:stack2}, begins few cycles prior to the maximum of the amplitude envelope $\sqrt{A_{H}^2(t) + A_{L}^2(t)}$ and includes the final cycles of the ringdown. 
As briefly discussed in the introduction, limiting the synchronization to this time window allows us to capture the high field dynamics of the binary black holes evolution.
It is impractical to synchronize the entire gravtiational wave emission of multiple BBH systems without using a transformation with a higher dimensional parameter space.

To obtain the optimal transformation parameters $\Omega_{12}$, we define the optimization functional $\Lambda(\Omega_{12})$, which is constructed to maximize the overlap of the reconstructed waveforms from the two events.
The functional is constructed for both detectors, however, first we 
introduce it for a single detector:

\begin{equation}
\label{eq:sigmaH1}
\sigma = \frac{\sum_{t = t_{i}}^{t_{f}} (h_1(t) - h_2(t))^2 + (H_1(t) - H_2(t))^2}{ \sum_{t = t_{i}}^{t_{f}} \sqrt{(h_1(t))^2 + (H_1(t))^2} \sqrt{(h_2(t))^2 + (H_2(t))^2} }\, ,
\end{equation} 
where $t_{i}$ and $t_{f}$ define the time window of optimization.
This statistic is calculated separately using Hanford  ($\sigma_{H}$) and Livingston ($\sigma_{L}$) waveforms.
Taking into account the waveforms from both detectors, the complete functional is:
\begin{equation}
\label{eq:sigma}
\Lambda(\Omega_{12}) = \frac{\sqrt{(\sigma_{H})^2 + (\sigma_{L})^2}}{\sqrt{2}}\, .
\end{equation}

Utilizing the optimzation program MINUIT~\cite{bib:minuit}, the overlap functional $\Lambda$ is minimized to determine the optimal transformation parameters ($\Omega_{12}$).
%The values of parameters $\alpha$, $\Delta t$, and $\Delta f$ are numerically determined.
The values of the $\alpha$, $\Delta t$, and $\Delta f$ parameters are initialized by aligning the reference times of the two BBH events $\tau_{1} = \tau_{2}$, and the corresponding frequencies of the signals at these times.
These initial values are fed into MINUIT where the optimal parameter values are determined.

%The initial values of the $\alpha$, $\Delta t$, and $\Delta f$ parameters are fed into MINUIT such that the reference times of the two BBH events are equal $\tau_{1} = \tau_{2}$, and the corresponding frequencies at these times are also equal $\eta_{1} = \eta_{2}$. 

In addition, the optimal parameters $\gamma_{H}$ and $\gamma_{L}$ are determined analytically.
Using the definition of the phase shift $\gamma$ given in Equation~\ref{eq:phase_shift} and the definition of the overlap functional for a single detector given in Equation~\ref{eq:sigmaH1}, we find an analytical solution for $\gamma$ from the equation $d \sigma/ d \gamma = 0$:  
\begin{equation} 
\label{eq:optimal_gamma}
\gamma = \tan^{-1}\Bigg(\frac{\sum_{t = t_{i}}^{t_{f}} h_1(t)H_2(t) - H_1(t)h_2(t)}{\sum_{t = t_{i}}^{t_{f}} h_1(t)h_2(t) + H_1(t)H_2(t)} \Bigg)\, .
\end{equation}
%This gives $\gamma_{H}$ for Hanford reconstructed waveforms and $\gamma_{L}$ for Livingston reconstructed waveforms.
The phase shifts $\gamma_{L}$ and $\gamma_{H}$ depend on the three other transformation parameters: $\gamma_{H,L} \equiv \gamma_{H,L}(\alpha, \Delta t, \Delta f)$.

Overall, the MINUIT optimization is performed with only three independent parameters: $%\Lambda \equiv
\Lambda(\alpha, \Delta t, \Delta f, \gamma_{H,L}(\alpha, \Delta t, \Delta f))$.
After the optimal transformation is determined,
%when $\Lambda$ reaches a minimum value.
the two synchronized events are then stacked together, producing the combined signal. Figure~\ref{fig:stack2} (bottom panels) shows the application of the synchronization method to match two reconstructed BBH events. 

\subsection{Organization of stacking for $N>2$ signals} 
\label{sec:StackingforMany}

\begin{figure}[htbp] % htbp stands for "here, top, bottom, page" 
\includegraphics[keepaspectratio=true,width=\linewidth]{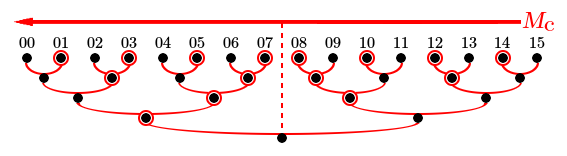}
\caption{
Binary tree for coherent stacking of 16 BBH events.
In the zeroth iteration, the BBH events, each denoted by a black dot, are ordered by chirp mass $M_c$ and matched into pairs. 
For each pair, the signal which lies closest to the center of the chirp mass spectrum is chosen as the template signal, denoted by a red circle.
The synchronization/summation method is applied to each pair, and we are left with 8 combined signals.
The iterative process continues until all of the events have been coherently stacked into one accumulated signal.
}
\label{fig:binarytree}
\end{figure}

Signals with similar astrophysical parameters, or similar signal morphology, are easier to synchronize.
To improve the synchronization for the entire population of selected BBH events, we order signals by the chirp mass $M_c$ as estimated by cWB, using $M_c$ as a measure of the morphology of a given signal.
The signals are matched into pairs and coherently stacked together in consecutive iterative steps as shown in Figure~\ref{fig:binarytree}. A stacking transformation is determined for each iterative step. 
When the iterative process concludes, a cumulative signal containing information about the mode structure for the BBH population is obtained. The iterative process is easily be expanded for any number of BBH events. 

We introduce the root-sum-squared amplitude to measure the efficiency of the coherent stacking procedure applied to NR injections:
\begin{equation}
\label{eq:h_tilde}
\tilde{h} = \sqrt{\sum\nolimits_{t = t_{i}}^{t_{f}} (h^2(t) + H^2(t)) \,dt}\, ,
\end{equation}
where $t_{i}$ and $t_{f}$ define the time window of optimization.
This amplitude is calculated separately for the reconstructed Hanford waveform ($\tilde{h}_{H}$) and the reconstructed Livingston waveform ($\tilde{h}_{L}$).
The efficiency of stacking multiple BBH signals is defined as:
\begin{equation}
\label{eq:efficiency}
\varepsilon = \frac{\tilde{h}_{sum}}{\sum_{n=1}^{N}\tilde{h}_{n}}\, ,
\end{equation}
%\begin{equation}
%\label{eq:efficiency}
%\varepsilon = \frac{{h}_{rss,sum}}{\sum_{i=0}^{N}{h}_{rss,i}} ,
%\end{equation}
where the numerator is the root-sum-squared amplitude of the stacked waveform, and the denominator is the sum of root-sum-squared amplitudes of the $N$ individual signals.
The stacking efficiency is calculated separately for Hanford waveforms ($\varepsilon_{H}$) and Livingston waveforms ($\varepsilon_{L}$). Furthermore, the stacking efficiency can be calculated both for the entire simulated signal or individual modes. This provides an estimation of the performance of our stacking procedure.

\section{Results}
\label{sec:Results}

The primary goal of the proposed stacking procedure is extracting information about the harmonic mode structure for a population of BBHs.
In particular, we are interested to determine: (i) if the sub-dominant modes of the BBH events are being coherently stacked, and (ii) if we can extract the sub-dominant modes from noisy data without \textit{a priori} knowledge about the signal model.

Using the coherent summation method, we stack the 16 BBH events listed in Table~\ref{tab:injections}.
As mentioned in Section~\ref{sec:MethodDescription}, the transformations used in the synchonization method depend only on the waveforms reconstructed by cWB, which do not rely on the signal models.

\begin{figure*}[!htb] 
\centering
 \makebox[\textwidth]{\includegraphics[width=.9\paperwidth]{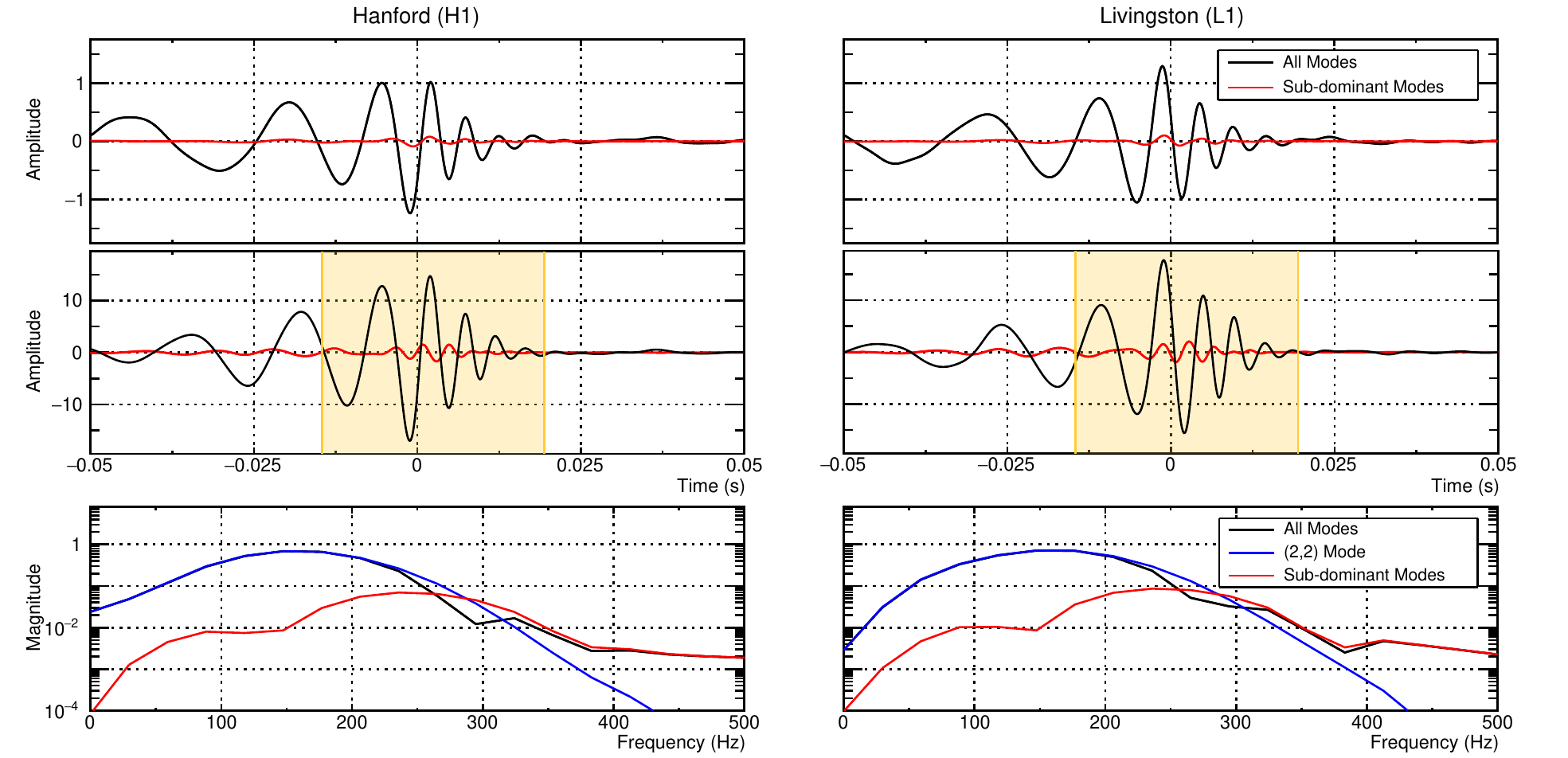}}
\caption{
Coherent stacking method applied to 16 BBH numerical relativity waveforms (\textit{left:} Hanford injections, \textit{right:} Livingston injections).
The synchronization transformations are determined explicitly by the reconstructed waveforms.
\textit{Top row:}
time series of one simulated BBH signal (Event 07 from Table~\ref{tab:injections}), shown for reference.
\textit{Middle row:}
16 simulated BBH events synchronized and summed. The amplitude of the combined signal is amplified by approximately a factor of 16 with respect to one BBH signal. The time window considered for optimization is shown in yellow.
\textit{Bottom row:}
frequency domain representation of the summed signal, produced with the segment of the waveform which lies in the optimization time window. The high frequency emission of the combined event is driven by the sub-dominant modes.
}
\label{fig:inj_analysis}
\end{figure*}

To test the coherence of the signal stacking procedure, we apply the stacking transformations directly to the numerical relativity waveforms of the corresponding BBH events.
In the time domain, the amplitude of the combined waveform is increased approximately by a factor of 16 with respect to a single numerical relativity waveform (top and middle panels, Figure~\ref{fig:inj_analysis}).
We expect this increase for coherently stacking 16 waveforms of similar amplitudes since the waveforms contain no detector noise. 
%In the frequency domain (bottom panels), the gravitational wave emission is dominanted by the $(2,2)$ mode.
In the frequency domain (bottom panel, Figure~\ref{fig:inj_analysis}), the gravitational wave emission is dominated at low freqencies by the (2,2) mode.
At higher frequencies, $\gtrsim 300\,$Hz, the emission is driven by the sub-dominant modes.
This behavior is consistent with the gravitational wave emission expected for a single BBH event.
These results show that the mode structure is preserved for the combined signal and that the sub-dominant modes are coherently stacked.

\begin{table}[ht]
\caption{
Each injected harmonic mode with its associated stacking efficiency per detector as defined in Equation~\ref{eq:efficiency}.
}
\begin{center}
\begin{tabular}{ c  c  c }
\hline \hline
Harmonic Mode & Efficiency $\varepsilon_{H}$ ($\%$) & Efficiency $\varepsilon_{L}$ ($\%$) \TBstrut\\  \hline 
$(2,2)$ & 97.6 & 97.7 \Tstrut\\
$(3,3)$ & 51.1 & 55.0 \\
$(4,4)$ & 45.0 & 47.2 \\
Sub-dominant & 47.3 & 50.8 \Bstrut\\ \hline 
\hline
\end{tabular}
\end{center}
\label{tab:efficiency}
\end{table}
In Table~\ref{tab:efficiency}, we calculate the efficiency of stacking given by Equation~\ref{eq:efficiency} for several harmonic modes to make a more definitive statement about the coherence of the stacked modes.
%In addition, we consider the sub-dominant modes to consist of the $(3,3)$, $(4,4)$, and $(5,5)$ harmonic modes and calculate the stacking efficiency.
We observe that the $(2,2)$ mode is the most efficiently stacked harmonic mode, consistent with the expected result since the gravitational wave emisison for BBHs is dominated by the $(2,2)$ mode in most orientations.
The sub-dominant modes are also seen to be coherently stacked, though less effectively.
The decrease in stacking efficiency for sub-dominant modes is partly  attributed to the different mode structure for each BBH event. 
As an example, for BBH systems with equal mass ratio $q=1$, the next leading order mode, other than the $(2,2)$ mode, is the $(4,4)$ mode. However, for BBH systems with $q>1$ the next leading order mode is the $(3,3)$ mode in most orientations. 
Overall, with the evidence for efficient stacking presented in Table~\ref{tab:efficiency}, and the predicted mode structure of a BBH system present in the combined simulated signal shown in Figure \ref{fig:inj_analysis}, we conclude that the sub-dominant modes are coherently stacked for the population of BBH events.

\begin{figure}[htpb] % htbp stands for "here, top, bottom, page" 
\includegraphics[keepaspectratio=true,width=\linewidth]
{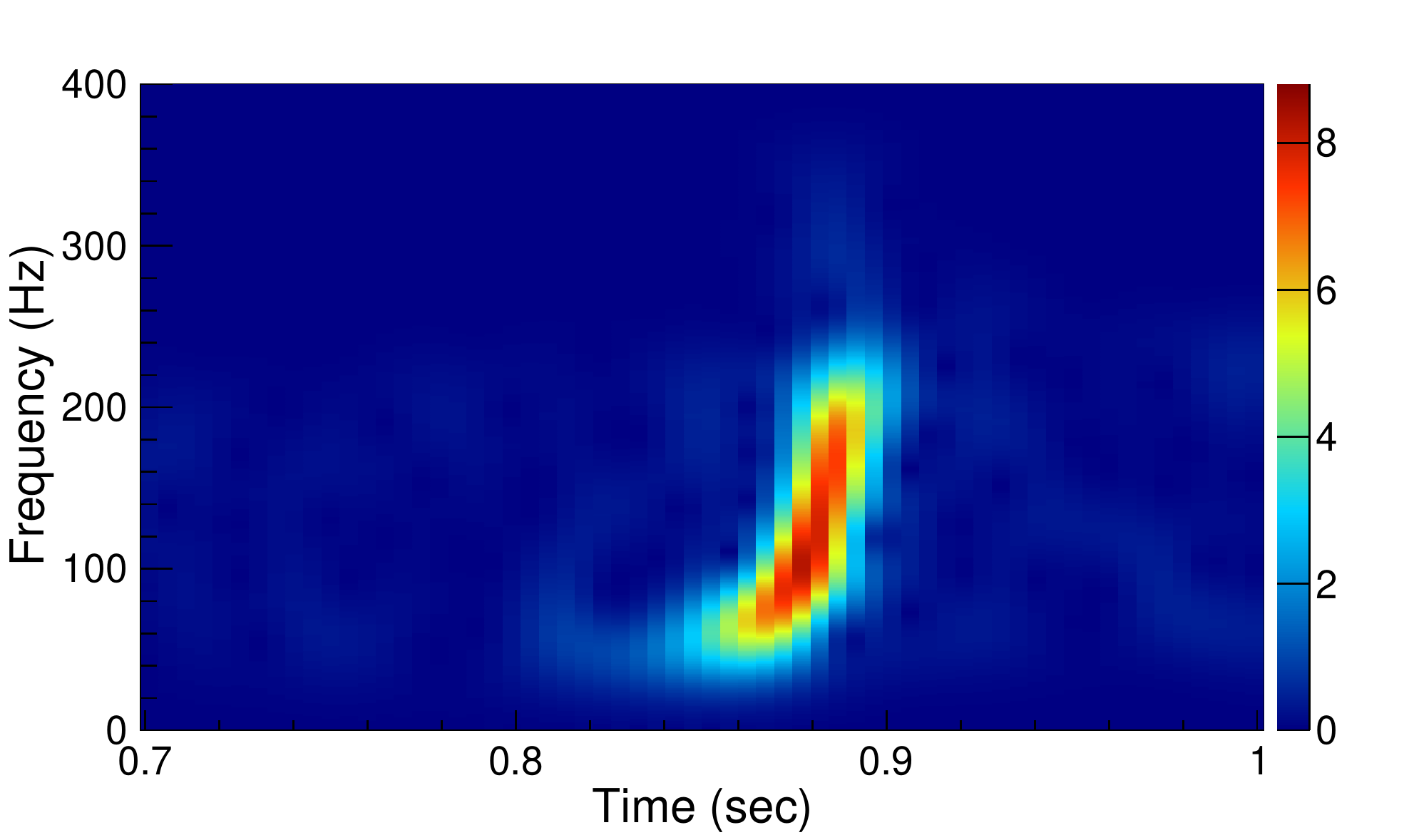}
\caption{
The signal is an accumulation of 16 BBH simulated events coherently stacked together. Time-frequency representation of cWB reconstruction in H1 for the summed signal. More high frequency behavior is captured compared to the reconstruction of a single BBH event.
}
\label{fig:tf_map_stacked}
\end{figure}

To make a measurement of sub-dominant modes in noisy data, we apply the synchronization procedure directly to the corresponding strain data which contain the detected BBH signals.
As before, the 16 BBH signals are synchronized by using the transformation set determined by the cWB reconstructed waveforms.
The 16 segments of strain data are then stacked to create a single data segment containing an combined BBH event.
The stacked data segment is processed again with the cWB pipeline, and the combined BBH event is recovered and reconstructed.
Figure~\ref{fig:tf_map_stacked} shows the time-frequency representation of the stacked waveform in the H1 detector. 

\begin{figure*}[!ptb] 
\centering
 \makebox[\textwidth]{\includegraphics[width=.9\paperwidth]{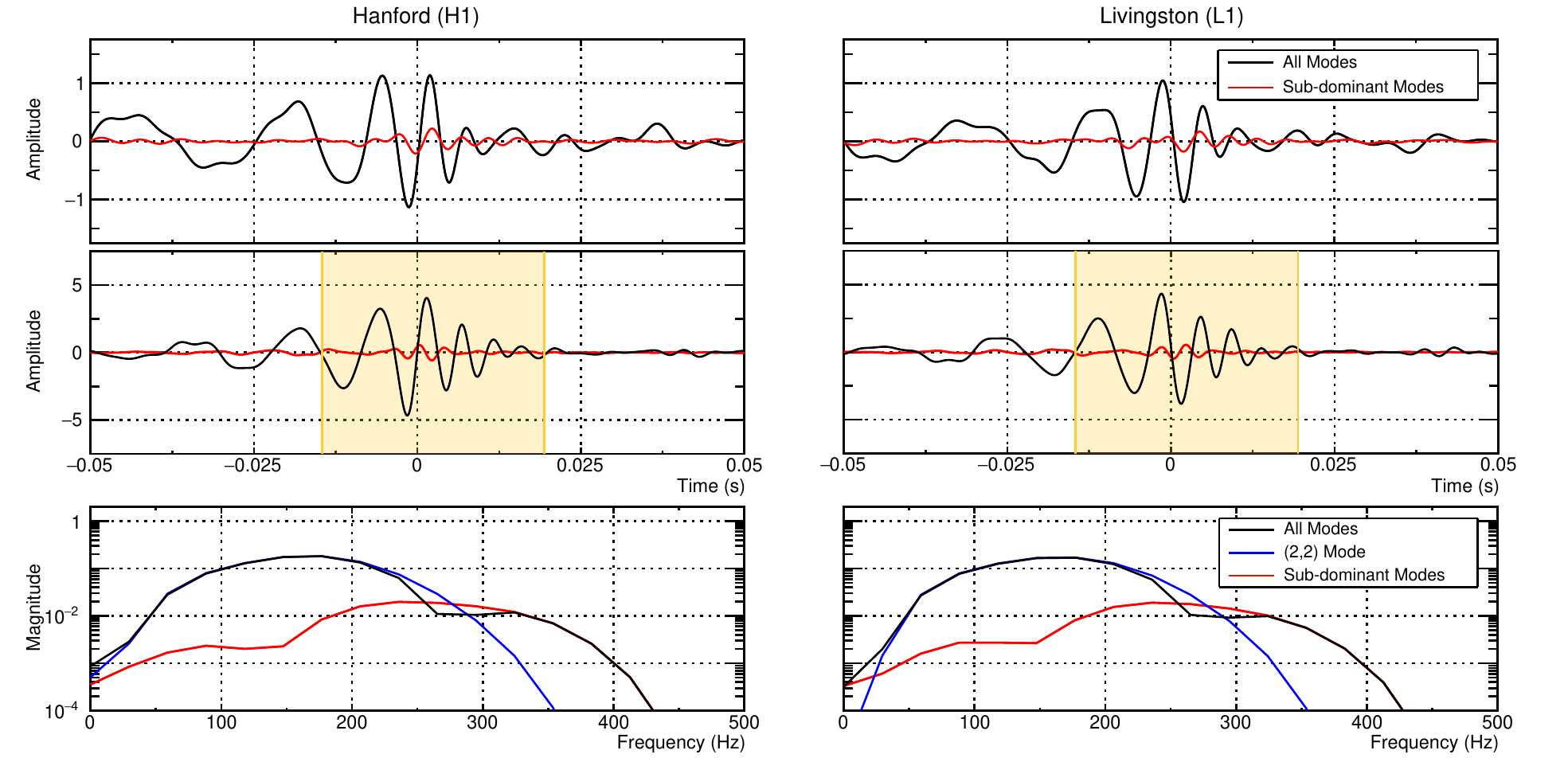}}
\caption{
Coherent stacking method applied to LIGO strain data with numerical relativity waveforms added. The cWB pipeline is used to recover and reconstruct the summed waveform.
The reconstructed sub-dominant modes waveform is produced by subtracting the (2,2) mode reconstruction from the reconstruction containing all harmonic modes. (\textit{left:} Hanford reconstructed waveofmrs, \textit{right:} Livingston reconstructed waveforms). \textit{Top row:} time series of numerical relativity waveform (Event 07 from Table~\ref{tab:injections}) injected into LIGO noise and reconstructed with cWB, shown for reference. \textit{Middle row:} time series of reconstructed summed signal. The amplitude of the reconstructed combined waveform is amplified by approximately a factor of 4 with respect to one BBH reconstructed signal.
The time window considered for optimization is shown in yellow. \textit{Bottom row:} frequency domain representation of the summed signal, produced with the segment of the waveform which lies in the optimization time window. The recovered high frequency emission of the combined event is driven by the sub-dominant modes.
}
\label{fig:recon_analysis}
\end{figure*}

The combined reconstructed waveforms are plotted in the time domain and Fourier domain in Figure~\ref{fig:recon_analysis}.
Compared to a reconstructed waveform for a single BBH event (top panel, Figure~\ref{fig:recon_analysis}), the amplitude of the combined reconstructed waveform is amplified approximately by a factor of $4$ (middle panel, Figure~\ref{fig:recon_analysis}).
This amplification is expected for coherent summation of 16 signals of similar amplitude since the simulated waveforms are added to LIGO detector noise.

The reconstructed aggregate signal also exhibits high frequency GW emission.
To confirm whether the additional captured emission is due to the sub-dominant modes, we applied the synchronization transformations to the same noisy signal models containing only the dominant $(2,2)$ mode.
The combined $(2,2)$ waveform reconstructed by cWB is depicted in blue in Figure~\ref{fig:recon_analysis}.
The residual,  depicted in red in Figure~\ref{fig:recon_analysis}, is produced by subtracting the reconstructed $(2,2)$ mode waveform from the reconstructed waveform containing all modes. This residual waveform is the contribution due to the sub-dominant modes for the combined BBH event.

In the frequency domain (bottom panel, Figure~\ref{fig:recon_analysis}), we observe that the sub-dominant modes are detected above the threshold of noise and reconstructed by cWB.
Furthermore, it is shown that the high frequency GW emission, $\gtrsim 300\,$Hz, is driven by the sub-dominant modes.
This result is in agreement with the high frequency emission of the combined numerical relativity waveform which is shown in Figure~\ref{fig:inj_analysis}. Therefore, the sub-dominant modes can be successfully extracted from the combined signal.
Here, we note that this combined signal contains all harmonic modes which will be distinguished from one another with a method not presented in this paper, which is under development.

%Although the figures may appear to convey similar information, each figure provides complementary information imperative to the aims of this study and the future of BBH mode structure research.
In summary, we conclude from Figure~\ref{fig:inj_analysis} that the sub-dominant modes for a population of BBHs are coherently stacked and we expect to resolve them.
In Figure~\ref{fig:recon_analysis}, we demonstrate that we are able to extract the sub-dominant modes from LIGO detector noise without the use of \textit{a priori} knowledge about the signal waveform.

\section{Conclusion}
\label{sec:Conclusion}

In this paper, we introduce a method for synchronizing and coherently stacking multiple BBH events. 
This generic synchronization procedure does not exploit information about the gravitational wave signal model.
Instead, the cWB pipeline is used to reconstruct unmodeled waveforms for each BBH event, and the stacking transformations are determined by  maximizing the overlap of reconstructed waveforms corresponding to different BBH events.

We demonstrate the efficient stacking of the (2,2) and sub-dominant harmonic modes for a population of 16 simulated BBH events. Also we demonstrate the possibility to extract the sub-dominant modes from LIGO detector noise without the use of \textit{a priori} knowledge of the signal model.

Future work will be dedicated to applying this method to the BBH events which have been detected by LIGO.
By comparing the results to the signal models, we can identify if the GW emission for a population of BBHs agrees with the predictions made by GR.
The end goal of event stacking is the measurement of the QNM spectrum which can accurately determine if the population of remnant black holes is described by the Kerr metric.

\newpage

%##################################################################################################
\section{Acknowledgments}
This work has been supported by NSF grant PHY 1806165.
The authors thank the California Institute of Technology and the Albert Einstein Institute for providing computational resources.
We would like to acknowledge with much appreciation F. Salemi for his help with mode selection. Thanks also to G. Vedovato for his patience and help with a variety of questions regarding simulations. We thank Gayathri V. and S. Tiwari for help with numerical relativity injections.

\bibliography{./myBib.bib}{}

\begin{thebibliography}{29}%
\makeatletter
\providecommand \@ifxundefined [1]{%
 \@ifx{#1\undefined}
}%
\providecommand \@ifnum [1]{%
 \ifnum #1\expandafter \@firstoftwo
 \else \expandafter \@secondoftwo
 \fi
}%
\providecommand \@ifx [1]{%
 \ifx #1\expandafter \@firstoftwo
 \else \expandafter \@secondoftwo
 \fi
}%
\providecommand \natexlab [1]{#1}%
\providecommand \enquote  [1]{``#1''}%
\providecommand \bibnamefont  [1]{#1}%
\providecommand \bibfnamefont [1]{#1}%
\providecommand \citenamefont [1]{#1}%
\providecommand \href@noop [0]{\@secondoftwo}%
\providecommand \href [0]{\begingroup \@sanitize@url \@href}%
\providecommand \@href[1]{\@@startlink{#1}\@@href}%
\providecommand \@@href[1]{\endgroup#1\@@endlink}%
\providecommand \@sanitize@url [0]{\catcode `\\12\catcode `\$12\catcode
  `\&12\catcode `\#12\catcode `\^12\catcode `\_12\catcode `\%12\relax}%
\providecommand \@@startlink[1]{}%
\providecommand \@@endlink[0]{}%
\providecommand \url  [0]{\begingroup\@sanitize@url \@url }%
\providecommand \@url [1]{\endgroup\@href {#1}{\urlprefix }}%
\providecommand \urlprefix  [0]{URL }%
\providecommand \Eprint [0]{\href }%
\providecommand \doibase [0]{http://dx.doi.org/}%
\providecommand \selectlanguage [0]{\@gobble}%
\providecommand \bibinfo  [0]{\@secondoftwo}%
\providecommand \bibfield  [0]{\@secondoftwo}%
\providecommand \translation [1]{[#1]}%
\providecommand \BibitemOpen [0]{}%
\providecommand \bibitemStop [0]{}%
\providecommand \bibitemNoStop [0]{.\EOS\space}%
\providecommand \EOS [0]{\spacefactor3000\relax}%
\providecommand \BibitemShut  [1]{\csname bibitem#1\endcsname}%
\let\auto@bib@innerbib\@empty
%</preamble>
\bibitem [{\citenamefont {{B.~P.~Abbott {\it et al.} (LIGO Scientific
  Collaboration, Virgo Collaboration)}}(2016{\natexlab{a}})}]{bib:GW150914}%
  \BibitemOpen
  \bibfield  {author} {\bibinfo {author} {\bibnamefont {{B.~P.~Abbott {\it et
  al.} (LIGO Scientific Collaboration, Virgo Collaboration)}}},\ }\href
  {\doibase 10.1103/PhysRevLett.116.061102} {\bibfield  {journal} {\bibinfo
  {journal} {Phys. Rev. Lett.}\ }\textbf {\bibinfo {volume} {116}},\ \bibinfo
  {pages} {061102} (\bibinfo {year} {2016}{\natexlab{a}})}\BibitemShut
  {NoStop}%
\bibitem [{bib(2018{\natexlab{a}})}]{bib:catalog_paper}%
  \BibitemOpen
  \href@noop {} {\  (\bibinfo {year} {2018}{\natexlab{a}})},\ \Eprint
  {http://arxiv.org/abs/1811.12907} {arXiv:1811.12907 [astro-ph.HE]}
  \BibitemShut {NoStop}%
%%CITATION = ARXIV:1811.12907;%%
\bibitem [{bib(2018{\natexlab{b}})}]{bib:catalog_paper_2}%
  \BibitemOpen
  \href@noop {} {\  (\bibinfo {year} {2018}{\natexlab{b}})},\ \Eprint
  {http://arxiv.org/abs/1811.12940} {arXiv:1811.12940 [astro-ph.HE]}
  \BibitemShut {NoStop}%
%%CITATION = ARXIV:1811.12940;%%
\bibitem [{\citenamefont {{B.~P.~Abbott {\it et al.} (LIGO Scientific
  Collaboration, Virgo Collaboration)}}(2016{\natexlab{b}})}]{bib:GW151226}%
  \BibitemOpen
  \bibfield  {author} {\bibinfo {author} {\bibnamefont {{B.~P.~Abbott {\it et
  al.} (LIGO Scientific Collaboration, Virgo Collaboration)}}},\ }\href
  {\doibase 10.1103/PhysRevLett.116.241103} {\bibfield  {journal} {\bibinfo
  {journal} {Phys. Rev. Lett.}\ }\textbf {\bibinfo {volume} {116}},\ \bibinfo
  {pages} {241103} (\bibinfo {year} {2016}{\natexlab{b}})}\BibitemShut
  {NoStop}%
\bibitem [{\citenamefont {{B.~P.~Abbott {\it et al.} (LIGO Scientific
  Collaboration, Virgo Collaboration)}}(2017{\natexlab{a}})}]{bib:GW170104}%
  \BibitemOpen
  \bibfield  {author} {\bibinfo {author} {\bibnamefont {{B.~P.~Abbott {\it et
  al.} (LIGO Scientific Collaboration, Virgo Collaboration)}}} (\bibinfo
  {collaboration} {LIGO Scientific and Virgo Collaboration}),\ }\href {\doibase
  10.1103/PhysRevLett.118.221101} {\bibfield  {journal} {\bibinfo  {journal}
  {Phys. Rev. Lett.}\ }\textbf {\bibinfo {volume} {118}},\ \bibinfo {pages}
  {221101} (\bibinfo {year} {2017}{\natexlab{a}})}\BibitemShut {NoStop}%
\bibitem [{\citenamefont {{B.~P.~Abbott {\it et al.} (LIGO Scientific
  Collaboration, Virgo Collaboration)}}(2017{\natexlab{b}})}]{bib:GW170608}%
  \BibitemOpen
  \bibfield  {author} {\bibinfo {author} {\bibnamefont {{B.~P.~Abbott {\it et
  al.} (LIGO Scientific Collaboration, Virgo Collaboration)}}},\ }\href
  {http://stacks.iop.org/2041-8205/851/i=2/a=L35} {\bibfield  {journal}
  {\bibinfo  {journal} {The Astrophysical Journal Letters}\ }\textbf {\bibinfo
  {volume} {851}},\ \bibinfo {pages} {L35} (\bibinfo {year}
  {2017}{\natexlab{b}})}\BibitemShut {NoStop}%
\bibitem [{\citenamefont {{B.~P.~Abbott {\it et al.} (LIGO Scientific
  Collaboration, Virgo Collaboration)}}(2017{\natexlab{c}})}]{bib:GW170814}%
  \BibitemOpen
  \bibfield  {author} {\bibinfo {author} {\bibnamefont {{B.~P.~Abbott {\it et
  al.} (LIGO Scientific Collaboration, Virgo Collaboration)}}},\ }\href
  {\doibase 10.1103/PhysRevLett.119.141101} {\bibfield  {journal} {\bibinfo
  {journal} {Phys. Rev. Lett.}\ }\textbf {\bibinfo {volume} {119}},\ \bibinfo
  {pages} {141101} (\bibinfo {year} {2017}{\natexlab{c}})}\BibitemShut
  {NoStop}%
\bibitem [{\citenamefont {{B.~P.~Abbott {\it et al.} (LIGO Scientific
  Collaboration)}}(2015)}]{bib:aLIGO}%
  \BibitemOpen
  \bibfield  {author} {\bibinfo {author} {\bibnamefont {{B.~P.~Abbott {\it et
  al.} (LIGO Scientific Collaboration)}}},\ }\href
  {http://stacks.iop.org/0264-9381/32/i=7/a=074001} {\bibfield  {journal}
  {\bibinfo  {journal} {Classical and Quantum Gravity}\ }\textbf {\bibinfo
  {volume} {32}},\ \bibinfo {pages} {074001} (\bibinfo {year}
  {2015})}\BibitemShut {NoStop}%
\bibitem [{\citenamefont {{F.~Acernese {\it et al.} (Virgo
  Collaboration)}}(2015)}]{bib:Virgo}%
  \BibitemOpen
  \bibfield  {author} {\bibinfo {author} {\bibnamefont {{F.~Acernese {\it et
  al.} (Virgo Collaboration)}}},\ }\href
  {http://stacks.iop.org/0264-9381/32/i=2/a=024001} {\bibfield  {journal}
  {\bibinfo  {journal} {Classical and Quantum Gravity}\ }\textbf {\bibinfo
  {volume} {32}},\ \bibinfo {pages} {024001} (\bibinfo {year}
  {2015})}\BibitemShut {NoStop}%
\bibitem [{\citenamefont {Ruffini}\ and\ \citenamefont
  {Wheeler}(1971)}]{RuffiniWheeler71}%
  \BibitemOpen
  \bibfield  {author} {\bibinfo {author} {\bibfnamefont {R.}~\bibnamefont
  {Ruffini}}\ and\ \bibinfo {author} {\bibfnamefont {J.}~\bibnamefont
  {Wheeler}},\ }\href@noop {} {\bibfield  {journal} {\bibinfo  {journal}
  {Physics Today}\ } (\bibinfo {year} {1971})}\BibitemShut {NoStop}%
\bibitem [{\citenamefont {Alpat}\ \emph {et~al.}(2010)\citenamefont {Alpat}
  \emph {et~al.}}]{bib:AMSsilicium}%
  \BibitemOpen
  \bibfield  {author} {\bibinfo {author} {\bibfnamefont {B.}~\bibnamefont
  {Alpat}} \emph {et~al.},\ }\href@noop {} {\bibfield  {journal} {\bibinfo
  {journal} {Nucl. Instrum. Meth.}\ }\textbf {\bibinfo {volume} {A613}},\
  \bibinfo {pages} {207} (\bibinfo {year} {2010})}\BibitemShut {NoStop}%
\bibitem [{\citenamefont {Vishveshwara}(1970)}]{bib:damped_osc}%
  \BibitemOpen
  \bibfield  {author} {\bibinfo {author} {\bibfnamefont {C.~V.}\ \bibnamefont
  {Vishveshwara}},\ }\href {\doibase 10.1103/PhysRevD.1.2870} {\bibfield
  {journal} {\bibinfo  {journal} {Phys. Rev. D}\ }\textbf {\bibinfo {volume}
  {1}},\ \bibinfo {pages} {2870} (\bibinfo {year} {1970})}\BibitemShut
  {NoStop}%
\bibitem [{\citenamefont {Press}(1971)}]{bib:Press71}%
  \BibitemOpen
  \bibfield  {author} {\bibinfo {author} {\bibfnamefont {W.~H.}\ \bibnamefont
  {Press}},\ }\href@noop {} {\bibfield  {journal} {\bibinfo  {journal} {The
  Astrophysical Journal}\ }\textbf {\bibinfo {volume} {170}},\ \bibinfo {pages}
  {L105} (\bibinfo {year} {1971})}\BibitemShut {NoStop}%
\bibitem [{\citenamefont {Nollert}(1999)}]{bib:QNM_sound}%
  \BibitemOpen
  \bibfield  {author} {\bibinfo {author} {\bibfnamefont {H.-P.}\ \bibnamefont
  {Nollert}},\ }\href {http://stacks.iop.org/0264-9381/16/i=12/a=201}
  {\bibfield  {journal} {\bibinfo  {journal} {Classical and Quantum Gravity}\
  }\textbf {\bibinfo {volume} {16}},\ \bibinfo {pages} {R159} (\bibinfo {year}
  {1999})}\BibitemShut {NoStop}%
\bibitem [{\citenamefont {Detweiler}(1980)}]{bib:Detweiler}%
  \BibitemOpen
  \bibfield  {author} {\bibinfo {author} {\bibfnamefont {S.}~\bibnamefont
  {Detweiler}},\ }\href@noop {} {\bibfield  {journal} {\bibinfo  {journal}
  {Astrophysical Journal, Part 1}\ }\textbf {\bibinfo {volume} {239}} (\bibinfo
  {year} {1980})}\BibitemShut {NoStop}%
\bibitem [{\citenamefont {Dreyer}\ \emph {et~al.}(2004)\citenamefont {Dreyer},
  \citenamefont {Kelly}, \citenamefont {Krishnan}, \citenamefont {Finn},
  \citenamefont {Garrison},\ and\ \citenamefont
  {Lopez-Aleman}}]{bib:bh_spectroscopy_dreyer}%
  \BibitemOpen
  \bibfield  {author} {\bibinfo {author} {\bibfnamefont {O.}~\bibnamefont
  {Dreyer}}, \bibinfo {author} {\bibfnamefont {B.}~\bibnamefont {Kelly}},
  \bibinfo {author} {\bibfnamefont {B.}~\bibnamefont {Krishnan}}, \bibinfo
  {author} {\bibfnamefont {L.~S.}\ \bibnamefont {Finn}}, \bibinfo {author}
  {\bibfnamefont {D.}~\bibnamefont {Garrison}}, \ and\ \bibinfo {author}
  {\bibfnamefont {R.}~\bibnamefont {Lopez-Aleman}},\ }\href
  {http://stacks.iop.org/0264-9381/21/i=4/a=003} {\bibfield  {journal}
  {\bibinfo  {journal} {Classical and Quantum Gravity}\ }\textbf {\bibinfo
  {volume} {21}},\ \bibinfo {pages} {787} (\bibinfo {year} {2004})}\BibitemShut
  {NoStop}%
\bibitem [{\citenamefont {Berti}\ \emph {et~al.}(2006)\citenamefont {Berti},
  \citenamefont {Cardoso},\ and\ \citenamefont
  {Will}}]{bib:bh_spectroscopy_berti}%
  \BibitemOpen
  \bibfield  {author} {\bibinfo {author} {\bibfnamefont {E.}~\bibnamefont
  {Berti}}, \bibinfo {author} {\bibfnamefont {V.}~\bibnamefont {Cardoso}}, \
  and\ \bibinfo {author} {\bibfnamefont {C.~M.}\ \bibnamefont {Will}},\ }\href
  {\doibase 10.1103/PhysRevD.73.064030} {\bibfield  {journal} {\bibinfo
  {journal} {Phys. Rev. D}\ }\textbf {\bibinfo {volume} {73}},\ \bibinfo
  {pages} {064030} (\bibinfo {year} {2006})}\BibitemShut {NoStop}%
\bibitem [{\citenamefont {Caudill}\ \emph {et~al.}(2012)\citenamefont
  {Caudill}, \citenamefont {Field}, \citenamefont {Galley}, \citenamefont
  {Herrmann},\ and\ \citenamefont {Tiglio}}]{bib:caudill}%
  \BibitemOpen
  \bibfield  {author} {\bibinfo {author} {\bibfnamefont {S.}~\bibnamefont
  {Caudill}}, \bibinfo {author} {\bibfnamefont {S.~E.}\ \bibnamefont {Field}},
  \bibinfo {author} {\bibfnamefont {C.~R.}\ \bibnamefont {Galley}}, \bibinfo
  {author} {\bibfnamefont {F.}~\bibnamefont {Herrmann}}, \ and\ \bibinfo
  {author} {\bibfnamefont {M.}~\bibnamefont {Tiglio}},\ }\href
  {http://stacks.iop.org/0264-9381/29/i=9/a=095016} {\bibfield  {journal}
  {\bibinfo  {journal} {Classical and Quantum Gravity}\ }\textbf {\bibinfo
  {volume} {29}},\ \bibinfo {pages} {095016} (\bibinfo {year}
  {2012})}\BibitemShut {NoStop}%
\bibitem [{\citenamefont {Meidam}\ \emph {et~al.}(2014)\citenamefont {Meidam},
  \citenamefont {Agathos}, \citenamefont {Van Den~Broeck}, \citenamefont
  {Veitch},\ and\ \citenamefont {Sathyaprakash}}]{bib:no-hair_TIGER}%
  \BibitemOpen
  \bibfield  {author} {\bibinfo {author} {\bibfnamefont {J.}~\bibnamefont
  {Meidam}}, \bibinfo {author} {\bibfnamefont {M.}~\bibnamefont {Agathos}},
  \bibinfo {author} {\bibfnamefont {C.}~\bibnamefont {Van Den~Broeck}},
  \bibinfo {author} {\bibfnamefont {J.}~\bibnamefont {Veitch}}, \ and\ \bibinfo
  {author} {\bibfnamefont {B.~S.}\ \bibnamefont {Sathyaprakash}},\ }\href
  {\doibase 10.1103/PhysRevD.90.064009} {\bibfield  {journal} {\bibinfo
  {journal} {Phys. Rev. D}\ }\textbf {\bibinfo {volume} {90}},\ \bibinfo
  {pages} {064009} (\bibinfo {year} {2014})}\BibitemShut {NoStop}%
\bibitem [{\citenamefont {Dain}\ \emph {et~al.}(2008)\citenamefont {Dain},
  \citenamefont {Lousto},\ and\ \citenamefont {Zlochower}}]{bib:dain}%
  \BibitemOpen
  \bibfield  {author} {\bibinfo {author} {\bibfnamefont {S.}~\bibnamefont
  {Dain}}, \bibinfo {author} {\bibfnamefont {C.~O.}\ \bibnamefont {Lousto}}, \
  and\ \bibinfo {author} {\bibfnamefont {Y.}~\bibnamefont {Zlochower}},\ }\href
  {\doibase 10.1103/PhysRevD.78.024039} {\bibfield  {journal} {\bibinfo
  {journal} {Phys. Rev. D}\ }\textbf {\bibinfo {volume} {78}},\ \bibinfo
  {pages} {024039} (\bibinfo {year} {2008})}\BibitemShut {NoStop}%
\bibitem [{\citenamefont {London}(2018)}]{bib:ringdown_London}%
  \BibitemOpen
  \bibfield  {author} {\bibinfo {author} {\bibfnamefont {L.~T.}\ \bibnamefont
  {London}},\ }\href@noop {} {\  (\bibinfo {year} {2018})},\ \Eprint
  {http://arxiv.org/abs/1801.08208} {arXiv:1801.08208 [gr-qc]} \BibitemShut
  {NoStop}%
%%CITATION = ARXIV:1801.08208;%%
\bibitem [{\citenamefont {Yang}\ \emph {et~al.}(2017)\citenamefont {Yang},
  \citenamefont {Yagi}, \citenamefont {Blackman}, \citenamefont {Lehner},
  \citenamefont {Paschalidis}, \citenamefont {Pretorius},\ and\ \citenamefont
  {Yunes}}]{Yang_stacking}%
  \BibitemOpen
  \bibfield  {author} {\bibinfo {author} {\bibfnamefont {H.}~\bibnamefont
  {Yang}}, \bibinfo {author} {\bibfnamefont {K.}~\bibnamefont {Yagi}}, \bibinfo
  {author} {\bibfnamefont {J.}~\bibnamefont {Blackman}}, \bibinfo {author}
  {\bibfnamefont {L.}~\bibnamefont {Lehner}}, \bibinfo {author} {\bibfnamefont
  {V.}~\bibnamefont {Paschalidis}}, \bibinfo {author} {\bibfnamefont
  {F.}~\bibnamefont {Pretorius}}, \ and\ \bibinfo {author} {\bibfnamefont
  {N.}~\bibnamefont {Yunes}},\ }\href {\doibase 10.1103/PhysRevLett.118.161101}
  {\bibfield  {journal} {\bibinfo  {journal} {Phys. Rev. Lett.}\ }\textbf
  {\bibinfo {volume} {118}},\ \bibinfo {pages} {161101} (\bibinfo {year}
  {2017})}\BibitemShut {NoStop}%
\bibitem [{\citenamefont {Da~Silva~Costa}\ \emph {et~al.}(2018)\citenamefont
  {Da~Silva~Costa}, \citenamefont {Tiwari}, \citenamefont {Klimenko},\ and\
  \citenamefont {Salemi}}]{Filipe_stacking}%
  \BibitemOpen
  \bibfield  {author} {\bibinfo {author} {\bibfnamefont {C.~F.}\ \bibnamefont
  {Da~Silva~Costa}}, \bibinfo {author} {\bibfnamefont {S.}~\bibnamefont
  {Tiwari}}, \bibinfo {author} {\bibfnamefont {S.}~\bibnamefont {Klimenko}}, \
  and\ \bibinfo {author} {\bibfnamefont {F.}~\bibnamefont {Salemi}},\ }\href
  {\doibase 10.1103/PhysRevD.98.024052} {\bibfield  {journal} {\bibinfo
  {journal} {Phys. Rev. D}\ }\textbf {\bibinfo {volume} {98}},\ \bibinfo
  {pages} {024052} (\bibinfo {year} {2018})}\BibitemShut {NoStop}%
\bibitem [{SXS()}]{SXS:catalog}%
  \BibitemOpen
  \href@noop {} {}\bibinfo {howpublished}
  {\url{http://www.black-holes.org/waveforms}}\BibitemShut {NoStop}%
\bibitem [{\citenamefont {Mroue}\ \emph {et~al.}(2013)\citenamefont {Mroue}
  \emph {et~al.}}]{bib:sxs2}%
  \BibitemOpen
  \bibfield  {author} {\bibinfo {author} {\bibfnamefont {A.~H.}\ \bibnamefont
  {Mroue}} \emph {et~al.},\ }\href {\doibase 10.1103/PhysRevLett.111.241104}
  {\bibfield  {journal} {\bibinfo  {journal} {Phys. Rev. Lett.}\ }\textbf
  {\bibinfo {volume} {111}},\ \bibinfo {pages} {241104} (\bibinfo {year}
  {2013})},\ \Eprint {http://arxiv.org/abs/1304.6077} {arXiv:1304.6077 [gr-qc]}
  \BibitemShut {NoStop}%
%%CITATION = ARXIV:1304.6077;%%
\bibitem [{\citenamefont {Buchman}\ \emph {et~al.}(2012)\citenamefont
  {Buchman}, \citenamefont {Pfeiffer}, \citenamefont {Scheel},\ and\
  \citenamefont {Szil\'agyi}}]{bib:sxs3}%
  \BibitemOpen
  \bibfield  {author} {\bibinfo {author} {\bibfnamefont {L.~T.}\ \bibnamefont
  {Buchman}}, \bibinfo {author} {\bibfnamefont {H.~P.}\ \bibnamefont
  {Pfeiffer}}, \bibinfo {author} {\bibfnamefont {M.~A.}\ \bibnamefont
  {Scheel}}, \ and\ \bibinfo {author} {\bibfnamefont {B.}~\bibnamefont
  {Szil\'agyi}},\ }\href {\doibase 10.1103/PhysRevD.86.084033} {\bibfield
  {journal} {\bibinfo  {journal} {Phys. Rev. D}\ }\textbf {\bibinfo {volume}
  {86}},\ \bibinfo {pages} {084033} (\bibinfo {year} {2012})}\BibitemShut
  {NoStop}%
\bibitem [{\citenamefont {Klimenko}\ \emph {et~al.}(2005)\citenamefont
  {Klimenko}, \citenamefont {Mohanty}, \citenamefont {Rakhmanov},\ and\
  \citenamefont {Mitselmakher}}]{sergey2005}%
  \BibitemOpen
  \bibfield  {author} {\bibinfo {author} {\bibfnamefont {S.}~\bibnamefont
  {Klimenko}}, \bibinfo {author} {\bibfnamefont {S.}~\bibnamefont {Mohanty}},
  \bibinfo {author} {\bibfnamefont {M.}~\bibnamefont {Rakhmanov}}, \ and\
  \bibinfo {author} {\bibfnamefont {G.}~\bibnamefont {Mitselmakher}},\ }\href
  {\doibase 10.1103/PhysRevD.72.122002} {\bibfield  {journal} {\bibinfo
  {journal} {Phys. Rev. D}\ }\textbf {\bibinfo {volume} {72}},\ \bibinfo
  {pages} {122002} (\bibinfo {year} {2005})}\BibitemShut {NoStop}%
\bibitem [{\citenamefont {Klimenko}\ \emph {et~al.}(2016)\citenamefont
  {Klimenko}, \citenamefont {Vedovato}, \citenamefont {Drago}, \citenamefont
  {Salemi}, \citenamefont {Tiwari}, \citenamefont {Prodi}, \citenamefont
  {Lazzaro}, \citenamefont {Ackley}, \citenamefont {Tiwari}, \citenamefont
  {Da~Silva},\ and\ \citenamefont {Mitselmakher}}]{sergey2016}%
  \BibitemOpen
  \bibfield  {author} {\bibinfo {author} {\bibfnamefont {S.}~\bibnamefont
  {Klimenko}}, \bibinfo {author} {\bibfnamefont {G.}~\bibnamefont {Vedovato}},
  \bibinfo {author} {\bibfnamefont {M.}~\bibnamefont {Drago}}, \bibinfo
  {author} {\bibfnamefont {F.}~\bibnamefont {Salemi}}, \bibinfo {author}
  {\bibfnamefont {V.}~\bibnamefont {Tiwari}}, \bibinfo {author} {\bibfnamefont
  {G.~A.}\ \bibnamefont {Prodi}}, \bibinfo {author} {\bibfnamefont
  {C.}~\bibnamefont {Lazzaro}}, \bibinfo {author} {\bibfnamefont
  {K.}~\bibnamefont {Ackley}}, \bibinfo {author} {\bibfnamefont
  {S.}~\bibnamefont {Tiwari}}, \bibinfo {author} {\bibfnamefont {C.~F.}\
  \bibnamefont {Da~Silva}}, \ and\ \bibinfo {author} {\bibfnamefont
  {G.}~\bibnamefont {Mitselmakher}},\ }\href {\doibase
  10.1103/PhysRevD.93.042004} {\bibfield  {journal} {\bibinfo  {journal} {Phys.
  Rev. D}\ }\textbf {\bibinfo {volume} {93}},\ \bibinfo {pages} {042004}
  (\bibinfo {year} {2016})}\BibitemShut {NoStop}%
\bibitem [{\citenamefont {{\it et al.}}(1994)}]{bib:minuit}%
  \BibitemOpen
  \bibfield  {author} {\bibinfo {author} {\bibfnamefont {J.~F.}\ \bibnamefont
  {{\it et al.}}},\ }\href
  {https://cds.cern.ch/record/2296388/files/minuit.pdf?version=1} {\emph
  {\bibinfo {title} {MINUIT: Function Minimization and Error Analysis Reference
  Manual}}} (\bibinfo {year} {1994})\BibitemShut {NoStop}%
\end{thebibliography}%

\end{document}